\documentstyle[aps,epsf,psfig,twocolumn]{revtex}
\begin{document}
\draft
\flushbottom
\twocolumn[
\hsize\textwidth\columnwidth\hsize\csname @twocolumnfalse\endcsname

\title{  
Crossover from marginal Fermi liquid to Luttinger liquid behavior
in carbon nanotubes}
\author{S. Bellucci $^1$ and J. Gonz\'alez $^2$ \\}
\address{
        $^1$INFN-Laboratori Nazionali di Frascati,
        P. O. Box 13, 00044 Frascati, Italy\\
        and INFM-Dipartimento di Fisica,
        Universit\`a di Roma Tor Vergata, Via della Ricerca Scientifica 1,
        00133 Roma, Italy.\\
        $^2$Instituto de Estructura de la Materia.
        Consejo Superior de Investigaciones Cient{\'\i}ficas.
        Serrano 123, 28006 Madrid. Spain.}
\date{\today} 
\maketitle
\widetext
\begin{abstract}
We study graphene-based electron systems with long-range Coulomb
interaction by performing an analytic continuation in the number
of dimensions. We characterize in this way the crossover between
the marginal Fermi liquid behavior of a graphite layer and the
Luttinger liquid behavior at $D = 1$. 
The former persists for any dimension above $D = 1$.
However, the proximity to the $D = 1$ fixed-point strongly
influences the phenomenology of quasi-onedimensional systems,
giving rise to an effective power-law behavior of observables like
the density of states. This applies to nanotubes of large
radius, for which we predict a lower bound of the corresponding
exponent that turns out to be very close to the value measured
in multi-walled nanotubes.
\end{abstract}


]
\narrowtext
\tightenlines
Recently there has been much interest in the search of
unconventional electron behavior deviating from the Fermi liquid
picture\cite{unconv}. Besides this, the other paradigm that is
well-established on theoretical grounds is the Luttinger liquid
behavior of one-dimensional (1D) electron systems\cite{sol,hal}. 
There have been suggestions that this behavior could be extended 
to two-dimensional (2D) systems,
in the hope that it may explain some of the features of the
copper-oxide materials\cite{and}. 
However, at least for the
Luttinger model, the analytic continuation in the number $D$ of
dimensions has shown that the Luttinger liquid behavior is lost
as soon as one departs from $D = 1$\cite{ccm,arri}.

Several authors have also analyzed the possibility that singular
interactions could lead to the breakdown of the Fermi liquid
picture\cite{sing}. 
With regard to real low-dimensional systems, such as
carbon nanotubes, the main electron interaction comes actually
from the long-range Coulomb potential $V(|{\bf r}|) \sim 
1/|{\bf r}| $. This is also the case of the 2D
layers in graphite, which have a vanishing
density of states at the Fermi level. Quite remarkably, a
quasiparticle decay rate linear in energy has been measured
experimentally in graphite\cite{exp}, 
pointing at the marginal Fermi liquid behavior in such 2D
layers. Due to the singular Coulomb interaction, the imaginary
part of the electron self-energy in the 2D system 
behaves at weak $g$ coupling like $g^2 \omega $\cite{expl}. 
It is crucial,
though, the fact that the effective coupling scales at low
energy as $g \sim 1/\log (\omega )$. This prevents the
logarithmic suppression of the quasiparticle weight, which gets
corrected by terms of order $g^2 \log (\omega ) \sim 1/\log
(\omega )$\cite{marg}. 

In this letter we investigate whether the long-range Coulomb
interaction may lead to the breakdown of the Fermi liquid
behavior at any dimension between $D = 1$ and 2. The issue is
significant for the purpose of comparing with recent
experimental observations of power-law behavior of the tunneling
conductance in multi-walled nanotubes\cite{mwnt}. 
These are systems whose description lies between that of a pure
1D system and the 2D graphite layer. It turns out, for instance,
that the critical exponent measured for tunneling  
into the bulk of the multi-walled nanotubes is $\alpha \approx
0.3$. 
This value is close to the exponent found for the
single-walled nanotubes\cite{bock,yao}. However, it is much 
larger than expected by taking into account the reduction 
due to screening ($\sim 1/\sqrt{N}$) in a
wire with a large number $N$ of subbands, what points towards 
sensible effects of
the long-range Coulomb interaction in the system.

We develop the analytic continuation in the number of dimensions
having in mind the low-energy modes of metallic nanotubes,
which have linear branches crossing at the Fermi level. From 
this picture, we build at general dimension $D$ a manifold of 
linear branches in momentum space crossing at a given Fermi 
point. We consider the hamiltonian
\begin{eqnarray}
H & = &  v_F \int_0^{\Lambda } d p |{\bf p}|^{D-1}
     \int  \frac{d\Omega }{(2\pi )^D} \;
   \Psi^{+} ({\bf p}) \;     \mbox{\boldmath $\sigma 
  \cdot $} {\bf p}  \;     \Psi ({\bf p})   \nonumber   \\ 
   \lefteqn{  +   e^2 \int_0^{\Lambda } d p |{\bf p}|^{D-1}
     \int  \frac{d\Omega }{(2\pi )^D}  \;
   \rho ({\bf p})   \;   \frac{c(D)}{|{\bf p}|^{D-1}}  \;
          \rho (-{\bf p})  \;\;\;\;\;\; }
\label{ham}
\end{eqnarray}
where the $\sigma_i $ matrices are defined formally by 
$ \{ \sigma_i , \sigma_j \} = 2\delta_{ij}$. Here $\rho ({\bf p})$
are density operators made of the electron modes
$\Psi ({\bf p})$, and $ c(D)/|{\bf p}|^{D-1} $ corresponds to 
the Fourier transform of the Coulomb potential in 
dimension $D$. Its usual logarithmic dependence on $|{\bf p}|$ 
at $D = 1$ is obtained by taking the 1D limit with $ c(D) = 
\Gamma ((D-1)/2)/(2\sqrt{\pi})^{3-D}$.

The dispersion relation $\varepsilon ({\bf p}) = \pm |{\bf p}|$ 
is that of Dirac fermions, with a vanishing density of states at 
the Fermi level above $D = 1$.
This ensures that the Coulomb interaction remains
unscreened in the analytic continuation. At $D = 2$ we
recover the low-energy description of the electronic properties
of a graphite layer, dominated by the presence of
isolated Fermi points with conical dispersion relation at the
corners of the Brillouin zone\cite{graph}. 

In the above picture, we are neglecting interactions that mix
the two inequivalent Fermi points common to the low-energy
spectra of graphite layers and metallic nanotubes. In the 
latter, such interactions have been considered in Refs.
\onlinecite{kbf} and \onlinecite{eg}, with the result that they
have smaller relative strength ($\sim 0.1/N$, in terms of the 
number $N$ of subbands) and remain small down to extremely 
low energies. More recently, the question has been addressed 
regarding the interactions in the graphite layer, and it also 
turns out that phases with broken symmetry cannot be realized,
unless the system is doped about half-filling\cite{nos} or 
it is in a strong coupling regime\cite{khves}.

We will accomplish a self-consistent solution of the model by
looking for fixed-points of the renormalization group
transformations implemented by the reduction of the cutoff
$\Lambda $\cite{sh}. As usual,
the integration of high-energy modes at that scale leads to 
the cutoff dependence of the parameters in the low-energy 
effective theory. We will see that the Fermi velocity $v_F$ grows 
in general as the cutoff is reduced towards the Fermi point. On the
other hand, the electron charge $e$ stays constant as $\Lambda
\rightarrow 0$. This comes from the fact that the polarizability
$\Pi $ does not show any singular dependence on the high-energy
cutoff $\Lambda $ for $D < 3$. The polarizability is then given by 
\begin{equation}
\Pi ({\bf k}, \omega_k) = b(D) \frac{v_F^{2-D} {\bf k}^2}
 { | v_F^2 {\bf k}^2 - \omega_k^2 |^{(3-D)/2} }\; ,\;  
\end{equation}
where $b(D) = \frac{2}{ \sqrt{\pi} } \frac{ \Gamma ( (D+1)/2 )^2 
   \Gamma ( (3-D)/2 ) }{ (2\sqrt{\pi})^D \Gamma (D+1) }$.

The dependence of $v_F$ on the cutoff $\Lambda $ implies 
an incomplete cancellation between self-energy and 
vertex corrections to the polarizability.
The dressed polarizability depends therefore on the effective Fermi 
velocity $v_F (\Lambda )$. The renormalized value of $v_F$ is 
determined by fixing it self-consistently to the value obtained 
in the electron propagator $G$ corrected by the self-energy 
contribution
\begin{eqnarray}
\Sigma ({\bf k}, \omega_k)  & = &  - e^2 \int_0^{\Lambda } 
     d p |{\bf p}|^{D-1}  \int \frac{d\Omega }{(2\pi )^D}
    \int \frac{d \omega_p}{2\pi }     \nonumber   \\ 
 \lefteqn{   G ({\bf k} - {\bf p}, \omega_k - \omega_p) 
 \frac{-i}{ \frac{|{\bf p}|^{D-1}}{c(D)} + e^2  \Pi ({\bf p}, 
    \omega_p) } .    }
\label{selfe}
\end{eqnarray}
  
The fixed-points of the renormalization group in the limit
$\Lambda \rightarrow 0$ determine the
universality class to which the model belongs. At $D = 2$, we
are bound to obtain the low-energy fixed-point at vanishing
coupling of the model of Dirac fermions with Coulomb
interaction\cite{marg}.  
On the other hand, at $D = 1$ there has to be presumably 
a fixed-point corresponding to Luttinger liquid behavior. We note,
however, that no solution of the model has been obtained yet
without carrying dependence on the transverse scale needed to
define the 1D logarithmic potential. Our dimensional regularization
overcomes the problem of introducing such external parameter, which
prevents a proper scaling behavior of the model\cite{wang}.

At general $D$, the self-energy (\ref{selfe}) shows a logarithmic
dependence on the cutoff at small frequency $\omega_k$ and small
momentum ${\bf k}$. This is the signature of the renormalization 
of the electron field scale and the Fermi velocity. In the 
low-energy theory with high-energy modes integrated out, the
electron propagator becomes
\begin{eqnarray}
\frac{1}{G}  & = &  \frac{1}{G_0} - \Sigma  
     \approx  Z^{-1} ( \omega_k - v_F  
  \mbox{\boldmath $\sigma \cdot$}{\bf k}) \;\;\;\;\;\;\;\;\; 
        \;\;\;\;\;\;\;\;\;  \;\;\;\;\;\;\;\;\;    \nonumber  \\
 \lefteqn{   -  Z^{-1} f(D)
 \sum_{n=0}^{\infty} (-1)^n g^{n+1}    \left( 
   \frac{n(3-D)}{n(3-D)+2}  \omega_k    \right.  }   \nonumber   \\ 
 \lefteqn{ +  \left.  \left(1 - \frac{2}{D} \frac{n(3-D)+1}{n(3-D)+2} 
   \right)   v_F \mbox{\boldmath $\sigma \cdot$} {\bf k}   
      \right)  h_n (D)   \log (\Lambda ) ,  }
\label{prop}
\end{eqnarray}
where $g = b(D) c(D) e^2 / v_F $,
$f(D) = \frac{1}{ 2^D \pi^{(D+1)/2} \Gamma (D/2) b(D) }$,
$h_n (D) = \frac{ \Gamma (n(3-D)/2 + 1/2) }
 { \Gamma (n(3-D)/2 + 1) }$ .
The quantity $Z^{1/2}$ represents the scale of the bare electron
field compared to that of the renormalized electron field for 
which $G$ is computed.

The renormalized propagator $G$ must be cutoff-independent,
as it leads to observable quantities in the quantum 
theory. This condition is enforced by fixing the dependence of
the effective parameters $Z$ and $v_F$ on $\Lambda $ as more
states are integrated out from high-energy shells. We get
the differential renormalization group equations
\begin{eqnarray}
\Lambda \frac{d}{d \Lambda} \log Z (\Lambda )  & = & 
 -  f(D) \sum_{n=0}^{\infty} (-1)^n g^{n+1}  
     \;\;\;\;\;\;\;\;\;   \;\;\;\;\;\;\;\;\;    \nonumber  \\  
  \lefteqn{  \frac{n(3-D)}{n(3-D)+2}   h_n (D) ,   }
\label{zflow}        
\end{eqnarray}
\begin{eqnarray}
\Lambda \frac{d}{d \Lambda} v_F (\Lambda )  & = & 
  - v_F f(D) \sum_{n=0}^{\infty} (-1)^n g^{n+1} 
      \;\;\;\;\;\;\;\;\;  \;\;\;\;\;\;\;\;\;    \nonumber     \\
 \lefteqn{  \left( 1 - 
  \frac{1}{D} \frac{n(3-D)(2-D) + 2}{n(3-D) + 2}  \right) h_n (D) . }
\label{vflow}
\end{eqnarray}

At $D = 2$, the right-hand-side of these equations can be 
summed up to the functions that have been found previously 
in the renormalization of the graphite layer\cite{marg}. 
Furthermore, 
they also provide meaningful expressions in the 1D limit. 
At $D = 1$, the right-hand-side of Eq. (\ref{vflow}) 
vanishes identically as a function of the variable $g $.
Therefore, the 1D model has formally a line of fixed-points, as
it happens in the case of a short-range interaction. The scaling
of the electron wavefunction can be read from the
right-hand-side of Eq. (\ref{zflow}), which becomes 
$(2 + g)/(2\sqrt{1 + g}) - 1$ at $D = 1$. This coincides with
the anomalous dimension that is found in the solution of 
the Luttinger model, what provides an independent check of the 
renormalization group approach to the 1D system.   

\begin{figure}

\par
\centering
\epsfxsize= 6cm
\epsfysize= 6cm
\epsfbox{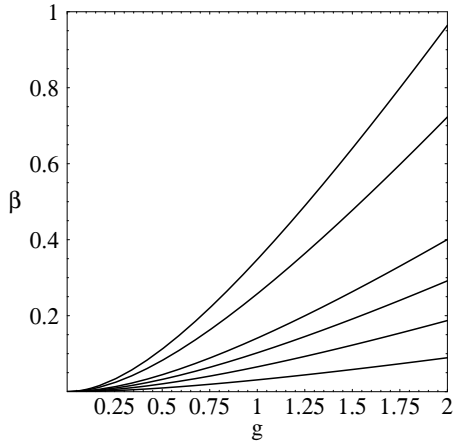}
\par

\caption{Plot of the $\beta$ function at the right-hand-side of 
Eq. (\ref{aflow}) for different dimensions. From top to bottom,
the curves correspond to the values $D = 2.0,1.7,1.4,1.3,1.2$
and $1.1$ .}
\label{one}   
\end{figure}

We have therefore a model that interpolates between marginal
Fermi liquid behavior, that is known to characterize the 2D model, and 
non-Fermi liquid behavior at $D = 1$. As the electron charge 
$e$ is not renormalized for $D < 3$, the scaling of
the effective coupling $g = b(D) c(D) e^2 / v_F $ is
given after Eq. (\ref{vflow}) by 
\begin{eqnarray}
\Lambda \frac{d}{d \Lambda} g (\Lambda )  & = & 
   f(D) \sum_{n=0}^{\infty} (-1)^n g^{n+2} 
   \;\;\;\;\;\;\;\;\;   \;\;\;\;\;\;\;\;\;   
             \;\;\;\;\;\;\;\;\;     \nonumber  \\
  \lefteqn{ \left( 1 - \frac{1}{D} 
   \frac{n(3-D)(2-D) + 2}{n(3-D) + 2}  \right)  h_n (D) .  }
\label{aflow}
\end{eqnarray}
The right-hand-side of Eq. (\ref{aflow}) is a monotonous
increasing function of $g $, for any dimension 
between 1 and 2, as observed in Fig. \ref{one} . 

\begin{figure}

\par
\centering
\epsfxsize= 8cm
\epsfysize= 8cm
\epsfbox{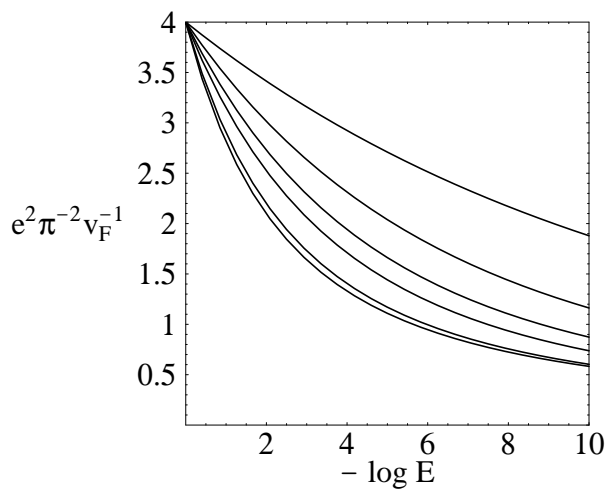}
\par

\caption{From top to bottom, curves of the effective coupling as
a function of the energy scale for $D = 1.1,1.2,1.3,1.4,1.7$    
and $2.0$ .}
\label{two}
\end{figure}

We find that, away from $D = 1$, there is only one fixed-point of
the renormalization group at  
$g = 0$. The scale dependence of the effective 
coupling $e^2 /v_F$ is displayed for different
values of $D$ in Fig. \ref{two}, where the flow to
the fixed-point is seen. Consequently,
the scale $Z$ of the wavefunction is not
renormalized to zero in the low-energy limit, and 
the quasiparticle weight remains finite above $D = 1$.
We conclude then that, even in a model that keeps the Coulomb 
interaction unscreened, the breakdown of the Fermi liquid 
behavior only takes place formally at $D = 1$.

The subtlety concerning the long-range Coulomb interaction
is that the function $c(D)$ diverges in the limit $D 
\rightarrow 1$. This is actually what transforms the power-law 
dependence of the potential into a logarithmic dependence at $D
= 1$. We observe that the 1D limit and the low-energy
limit $\Lambda \rightarrow 0$ do not commute. If we stick 
to $D = 1$, we obtain a divergent coupling $g $
for the Coulomb interaction as well as a divergent electron 
scaling dimension. At any dimension slightly above $D = 1$, 
however, the fixed-point is at $g = 0$, with its 
corresponding vanishing anomalous dimension.

In order to understand whether the 1D model has any stable
fixed-point for finite values of $e^2 /v_F$, one can  
study the model by performing an expansion in powers of 
$g^{-1}$. The Fermi velocity is renormalized by terms 
that are analytic near the point $g = \infty$, and that 
lead to the scaling equation 
\begin{equation}
\Lambda \frac{d}{d \Lambda} v_F (\Lambda )  =   \\ 
   v_F f(D) \left( 3 - D - \frac{2}{D} \right)
   \frac{ \Gamma (D/2 - 1) }{ \Gamma ((D+1)/2) }
\label{inf}
\end{equation}
up to terms of order  $O(g^{-1})$.
In the limit $D \rightarrow 1$, $g \rightarrow \infty$,
the right-hand-side of Eq. (\ref{inf}) vanishes identically.
This confirms, on nonperturbative grounds, that the 1D model
with the Coulomb interaction has  
a line of fixed-points covering all values of $e^2 /v_F$.

\begin{figure}

\par
\centering
\epsfxsize= 6cm
\epsfysize= 6cm
\epsfbox{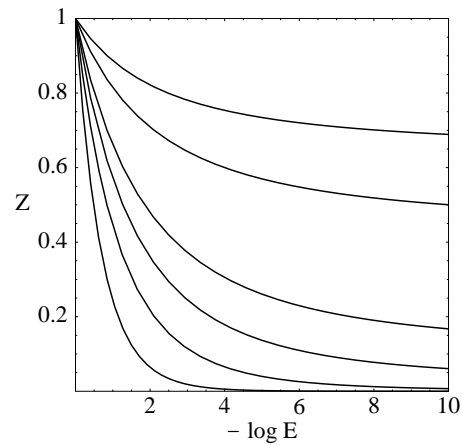}
\par

\caption{Plot of the electron field scale $Z$ for a value of the
bare coupling $e^2 /(\pi^2 v_F) = 4.0$. The different curves
correspond, from top to bottom, to $D = 2.0,1.7,1.4,1.3,1.2$
and $1.1$ .}
\label{three}    
\end{figure}

In the vicinity of $D = 1$, the presence of such critical
line becomes sensible, and a crossover takes place to a 
behavior with a sharp reduction of the quasiparticle 
weight. This can be seen in the renormalization of the electron
field scale $Z$, displayed in Fig. \ref{three}. For values 
of $D$ above $\approx 1.2$, we have a clear signature of
quasiparticles in the value of $Z$ at low energies. For lower
values of $D$, the picture cannot be distinguished from that 
of a vanishing quasiparticle weight for all practical purposes.
The drastic suppression of the electron field scale $Z$ takes
place over a variation of only two orders of magnitude in the 
energy scale.


\begin{figure}

\par
\centering
\epsfxsize= 6cm
\epsfysize= 6cm
\epsfbox{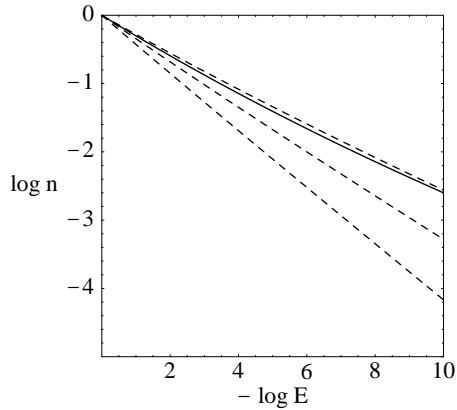}
\par

\caption{Plot of the logarithm of the density of states for a
value of the bare coupling $e^2 /(\pi^2 v_F) = 0.5$.
The solid line corresponds to $D = 1.1$ and the dashed lines 
correspond, from top to bottom, to $D = 1.2, 1.3$ and $1.4$ .}
\label{four}
\end{figure}

The above picture allows us to make contact with the
experiments carried out in multi-walled nanotubes\cite{mwnt}.
In the proximity of the $D = 1$ fixed-point, the density of
states displays an effective power-law behavior, 
with an increasingly large exponent. Moving to the other
side of the crossover, the density of states approaches the
well-known behavior of the graphite layer, $n( \varepsilon )
\sim |\varepsilon |$. In Fig. \ref{four} we give the 
representation of the density of states
\begin{equation}
n( \varepsilon ) \sim Z( \varepsilon ) |\varepsilon |^{D-1}
\end{equation}
for several dimensions approaching $D = 1$.

A value 
$e^2 /(\pi^2 v_F) = 0.5 $ for the bare coupling is appropriate
for typical multi-walled nanotubes, as it takes into account the
reduction due to the interaction with the inner metallic
cylinders\cite{egger}. We observe that the exponents of $n(
\varepsilon )$ at different dimensions are always larger than 
a lower bound $\alpha \approx 0.26$. This is in agreement with the  
values measured experimentally. 
Our analysis stresses the need of an appropriate description
of the dimensional crossover between one and two dimensions,
showing that the picture of a thick nanotube as an aggregate 
of 1D channels does not allow to obtain the correct values of
the critical exponents.

To summarize, we have studied the renormalization of the 
Coulomb interaction in graphene-based structures. 
We have made a rigorous characterization of the different 
behaviors, as we have proceeded by identifying the fixed-points 
of the theory. We have seen that the Fermi liquid behavior persists
formally for any dimension above $D = 1$, as it also happens in
the case of a short-range interaction\cite{ccm}. 
On the other hand, the proximity to the 1D
fixed-point influences strongly the phenomenology of 
real quasi-onedimensional systems, giving rise to an effective 
power-law behavior of observables like the tunneling density 
of states. This is the case of the multi-walled 
nanotubes, for which we predict a lower bound for the 
corresponding exponent that turns out to be very close to 
the value measured experimentally.

Financial support from CICyT (Spain)
and CAM (Madrid, Spain) through grants PB96/0875 and
07N/0045/98 is gratefully
acknowledged.

\end{document}